\documentclass[10pt,final,journal,twocolumn]{IEEEtran}
\usepackage{amsmath}
\usepackage{color,soul}
\usepackage{graphicx}
\usepackage{kantlipsum}
\usepackage{caption}
\usepackage[colorlinks,]{hyperref}
\usepackage[noadjust]{cite}

\begin{document}
	
%	\maketitle
	
	\title{Synthetic Elastography using B-mode Ultrasound through a Deep Fully-Convolutional Neural Network}
	
	\author{R. R. ~Wildeboer, R. J. G. ~van Sloun, C. K. ~Mannaerts, P. H. ~Moraes, G. ~Salomon, M.C. ~Chammas, H. ~Wijkstra, M. ~Mischi
	\thanks{R.R. Wildeboer, R.J.G van Sloun, H. Wijkstra and M. Mischi are with the Lab of Biomedical Diagnostics, Department of Electrical Engineering, Eindhoven University of Technology, The Netherlands. (e-mail: r.r.wildeboer@tue.nl; r.j.g.v.sloun@tue.nl)}
	\thanks{C.K. Mannaerts and H. Wijkstra are with the Department of Urology, Academic University Medical Centres, University of Amsterdam, The Netherlands.}
	\thanks{G. Salomon is with the Department of Urology, University Hospital Hamburg-Eppendorf, Germany.}
	\thanks{P. H. ~Moraes and M.C. Chammas are with the Department of Radiology, Universidade de S\~ao Paulo Faculdade de Medicina Hospital das Cl\'inicas.}
	\thanks{}
	\thanks{\copyright 2020 IEEE.  Personal use of this material is permitted.  Permission from IEEE must be obtained for all other uses, in any current or future media, including reprinting/republishing this material for advertising or promotional purposes, creating new collective works, for resale or redistribution to servers or lists, or reuse of any copyrighted component of this work in other works.}
	}
	
	\maketitle	
	
	\begin{abstract}
	Shear-wave elastography (SWE) permits local estimation of tissue elasticity, an important imaging marker in biomedicine. This recently-developed, advanced technique assesses the speed of a laterally-travelling shear wave after an acoustic radiation force ``push'' to estimate local Young's moduli in an operator-independent fashion. In this work, we show how synthetic SWE (sSWE) images can be generated based on conventional B-mode imaging through deep learning. Using side-by-side-view B-mode/SWE images collected in 50 patients with prostate cancer, we show that sSWE images with a pixel-wise mean absolute error of 4.5$\pm$0.96~kPa with regard to the original SWE can be generated. Visualization of high-level feature levels through t-Distributed Stochastic Neighbor Embedding reveals {substantial} overlap between data from {two} different scanners. {Qualitatively, we examined the use of the sSWE methodology for B-mode images obtained with a scanner without SWE functionality. We also examined the use of this type of network in elasticity imaging in the thyroid. Limitations of the technique reside in the fact that networks have to be retrained for different organs, and that the method requires standardization of the imaging settings and procedure. Future research will be aimed at development of sSWE as an elasticity-related tissue typing strategy that is solely based on B-mode ultrasound acquisition, and the examination of its clinical utility.}
	\end{abstract}


\section{Introduction}
Tissue elasticity is an important biomarker of cancer. Prostate cancer, for example, is characterized by increased stiffness \cite{Correas2013}, thyroid and liver nodules can be discriminated based on their elasticity \cite{Henry2010,Barr2018}, and also breast lesions are typically diagnosed based on their elastic properties \cite{Chang2011}. It is also increasingly used to image musculoskeletal pathologies in e.g. muscles, tendons, and ligaments \cite{Taljanovic2017}. Over the last few decades, this has spurred considerable advances in the development of elasticity imaging. 

Ultrasound-based elasticity imaging, that is, ultrasound elastography, has played a major role in these developments \cite{Sigrist2017}. So-called quasi-static elastographic {(QSE)} strain imaging allows for the relative assessment of tissue deformation due to externally applied stress, but as this stress is often manually delivered, the technique remains operator dependent and limited to superficial organs. Therefore, more recently, dynamic elastography techniques were developed where tissue deformation induced by an acoustic radiation force ``push'' pulse is quantified to obtain more objective and reproducible measures of elasticity \cite{Gennisson2013}. At this moment, we distinguish especially acoustic radiation force imaging (ARFI) and shear-wave elastography \cite{Nightingale2011,Taljanovic2017}. The first method analyses tissue displacement resulting from a “push” pulse along the beam path, whereas the latter relies on the speed of transversally-travelling shear waves to estimate tissue elasticity. The tissue elasticity is quantified by the Young's modulus, that is, the ratio between stress and strain. 

SWE requires advanced ultrafast acquisition schemes with frame rates of $\sim$1000~Hz to accurately assess tissue deformation and shear-wave dynamics \cite{Gennisson2013,Bouchet}. Moreover, ultrasound transducers have to be sufficiently equipped to allow for the generation of acoustic radiation force pulses as well as ultrafast imaging of the shear wave displacements \cite{Sarvazyan1998}. Although several techniques and sequences have been developed to enable SWE on commercial scanners, the frame rate of conventional B-mode ultrasound cannot be reached as it requires long settling times and multiple ``push'' pulses to reliably generate an elastogram. 

\begin{figure}[]
	\centering
	\includegraphics[trim=-0.2in 0in -0.2in 0in,width=0.5\textwidth]{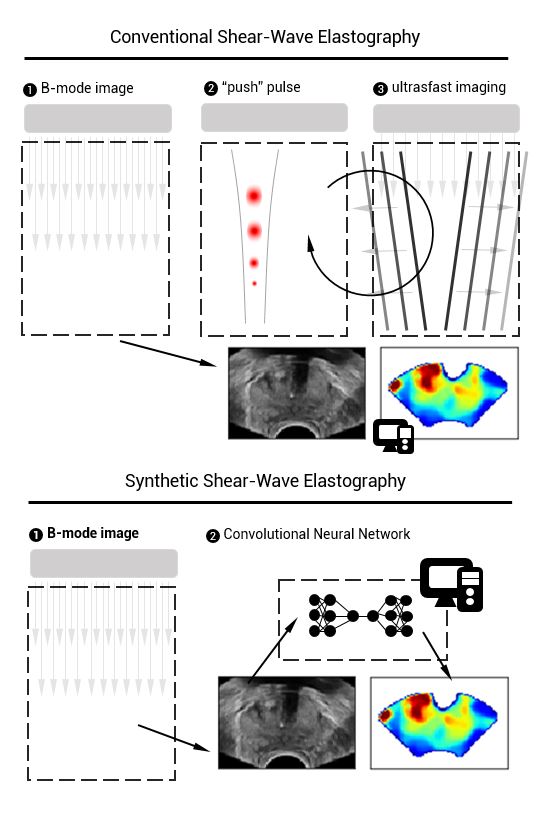}
	\caption{Schematic implementation of conventional SWE and synthetic SWE.}
	\label{idea}
\end{figure}

Realizing that conventional B-mode ultrasound assesses tissue echogenicity rather than tissue elasticity, we here propose that both properties can be expected to be linked through their dependence on the underlying tissue structure. In this work, we exploit this fact by designing a deep fully-convolutional neural network (DCNN) that is able to assess echogenic patterns in B-mode ultrasound that are useful for elasticity-related tissue typing (see Figure~\ref{idea}). Whereas deep-learning strategies were already proposed for estimation of speed of sound \cite{Feigin2018}, extraction of strain images from radio frequency data \cite{Wu2018,Kibria2018}, and for processing of conventional SWE sequences \cite{Ahmed2019}, we train our network to directly map B-mode ultrasound towards the corresponding elasticity images obtained through SWE.\\

\section{Materials and Methods}
\subsection{Data Acquisition}
At the Martini Clinic in Hamburg, supersonic shear imaging was performed in 50 patients that were diagnosed with prostate cancer using the Aixplorer ultrasound scanner (SuperSonic Imagine, Aix-en-Provence, France). At least 3 image planes (basal, mid-gland and apical orientation) were recorded per patient, defining regions of interest (ROIs) that covered the entire prostate and smaller ROIs that only covered one side or a suspicious area. {The acquired SWE images were obtained with minimal preload and such that a steady position was maintained for 5 s} \cite{Mannaerts2018a}{; the resulting images had a pixel spacing of 0.16 mm, a zoom of 120\%, a focus positioned above the anterior edge of the gland and a maximum elasticity scale of 70 kPa.}

At least 9 {and at most 15} images were obtained per patient. {In some cases, the operator chose to make additional SWE images of suspicious structures or repeat an acquisition to obtain a higher-quality image; to avoid introducing any bias by image selection, all images were used for training}. We extracted the Young's modulus data from the SWE acquisitions, {as well as the estimation confidence (ranging from 0 to 1) calculated by the machine to reflect the local elastogram quality.} Pre-processing involved alignment of the side-by-side B-mode and SWE data, {followed by physical regridding.} This entailed downsampling onto a conveniently-scaled 96$\times$64 grid from an original extracted size of 600$\times$400 pixels {by means of bilinear interpolation}. The B-mode images were subsequently {scaled from 0-255 to 0-1}. Likewise, the elastography data were scaled by 100 kPa so that clinically-relevant Young's moduli also scale from 0 to 1. 

{In addition, full-screen B-mode images {(sized 600$\times$900 with a pixel spacing of 0.095 mm, focus in the centre of the gland, and a zoom of 120\%)} were obtained in roughly the same imaging planes using the B-mode-only acquisition protocol. As these B-mode images are formed outside of the (side-by-side) SWE sequence, they exhibit a different resolution before regridding, slightly different image features, and any information leak from the SWE formation into the B-mode is prevented. These images were not used for training.}

In order to establish the use of sSWE in a device that does not feature SWE itself, B-mode and QSE recordings were performed in 10 patients at the Academic Medical Centre (University Hospital, Amsterdam) using an iU22 scanner (Philips Healthcare, Bothell, WA) equipped with a C10-3v probe. {The extracted images were 450$\times$300 pixels in size (i.e., downsample factor of 4.7), had a focus positioned in the middle of the prostate, and a pixel size of 0.12 mm.} QSE allows for the extraction of relative stiffness by assessment of tissue compression and decompression upon cyclic manual pressure asserted by the ultrasound operator \cite{Gennisson2013}. These quasi-static elastograms allow a qualitative evaluation of sSWE, {as these images reflect elasticity on a larger scale  than SWE's pixelwise quantification  due to the operator-dependency of QSE and its relative rather than absolute nature.}

In addition, we studied the applicability of sSWE in other organs by using SWE recordings of thyroid nodules. For this, we used a different dataset collected in 215 patients at the Hospital das Cl\'inicas da Faculdade de Medicina da Universidade in S\~ao Paulo, Brazil. The recordings were obtained using a LOGIQ E9 ultrasound device (GE HealthCare, Wauwasota, WI, USA) equipped with a 9L probe. Approximately three images per patient were collected, {with an extracted size of 480$\times$320 pixels (i.e., downsample factor of 5), pixel size of 0.12 mm, and focus at approximately 2 cm.} Here, the shear-wave speed was extracted from the images as measure for elasticity. Subsequently, the elastograms were subjected to normalisation (i.e., by 10 m/s, also to scale the clinically-relevant speeds between 0 and 1) and the same pre-processing procedure as the prostate recordings. The confidence of the shear-wave speed estimation {(comparable to the confidence measure available in the prostate recordings)} could not be extracted.

\begin{figure*}[]
	\centering
	\includegraphics[trim=0.75in 0.5in 0.75in 0.5in,width=0.9\textwidth]{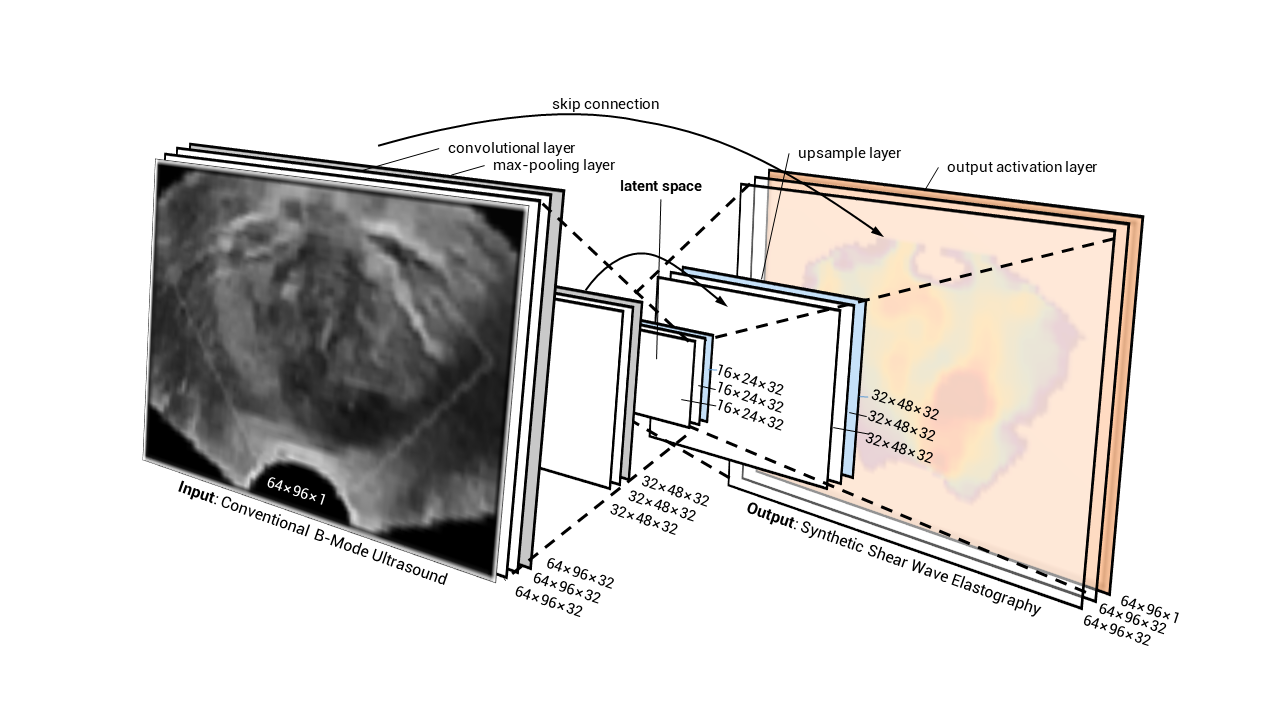}
	\caption{Schematic representation of the proposed DCNN architecture for the synthesis of shear-wave elastography from conventional B-mode ultrasound.}
	\label{architecture}
\end{figure*}

\subsection{Neural Network Architecture}
We designed a DCNN that serves as an end-to-end nonlinear mapping function transforming 2D B-mode ultrasound images to 2D synthetic SWE images. To this end, we employ {a U-net-like architecture} \cite{Ronneberger2015}{, featuring a general encoder-decoder shape} in which a hierarchy of features is consecutively extracted from the B-mode data to yield a latent feature space. These features are subsequently used to construct an SWE image by a decoding network that approximately mirrors the encoding part. This type of network has been used frequently for image segmentation and reconstruction tasks \cite{Badrinarayanan2017,Noh2015,Han2017}. The network was contained direct “skip” connections from the encoder filter layer to its equally-sized decoder counterpart, as introduced by \cite{Ronneberger2015}. By transferring the encoder layer output across the latent space and concatenating it to the larger-scale model features during decoding, we enable our network to combine fine and course level information and generate higher-resolution SWE estimations. See Figure~\ref{architecture} for an overview of the DCNN architecture.

The convolutional layers of the proposed network comprised a bank of 2D 3$\times$3-pixel convolutional filters (described by the filter weights) and biases of which the results were subsequently passed through a non-linear activation function. Every convolution layer maps its input to 32 feature maps. Leaky Rectified Linear Units (Leaky ReLUs; i.e., $f(x) = \mathrm{max}(\alpha~\cdot~x,x))$ with an $\alpha$-value of 0.1 were adopted as non-linear activation functions to minimize the risk of vanishing gradients \cite{Maas}.

\begin{figure*}[]
	\centering
	\includegraphics[trim=2.2in 1.2in 2in 1in,clip,width=0.98\textwidth]{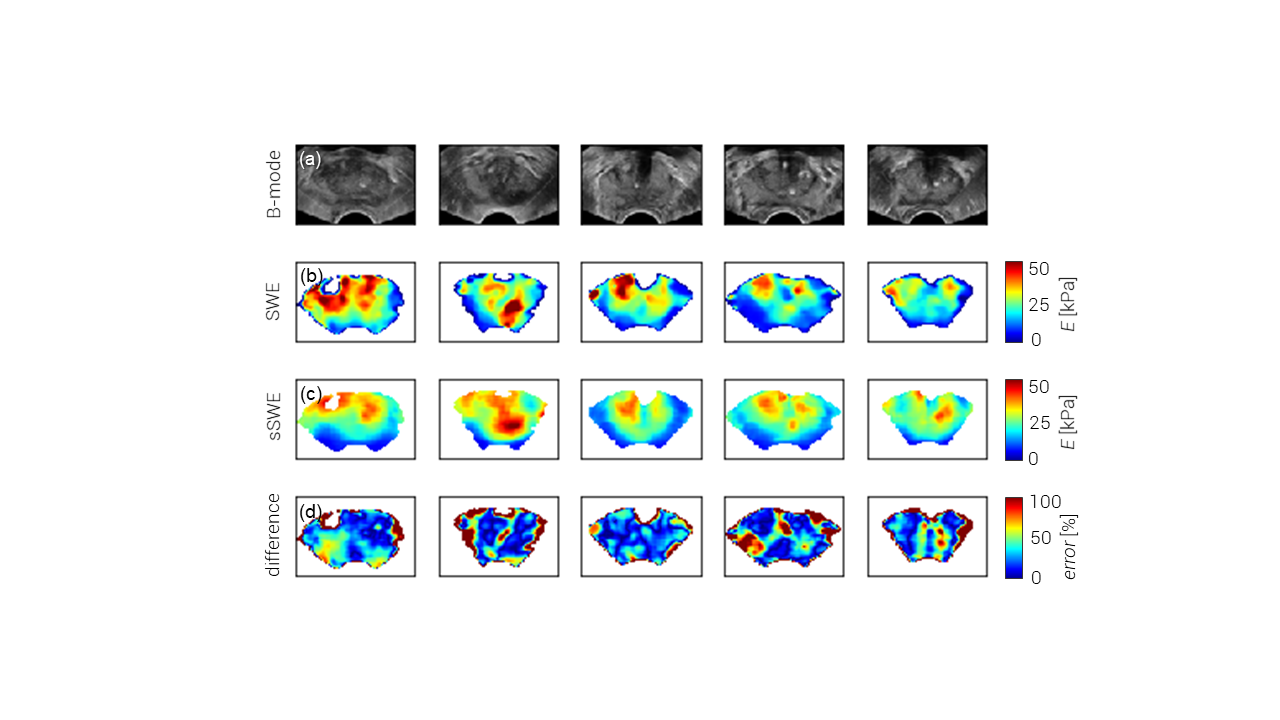}
	\label{examples}
\end{figure*}

\begin{figure*}[]
	\centering
	\includegraphics[trim=2.2in 1in 2in 1.5in,clip,width=0.98\textwidth]{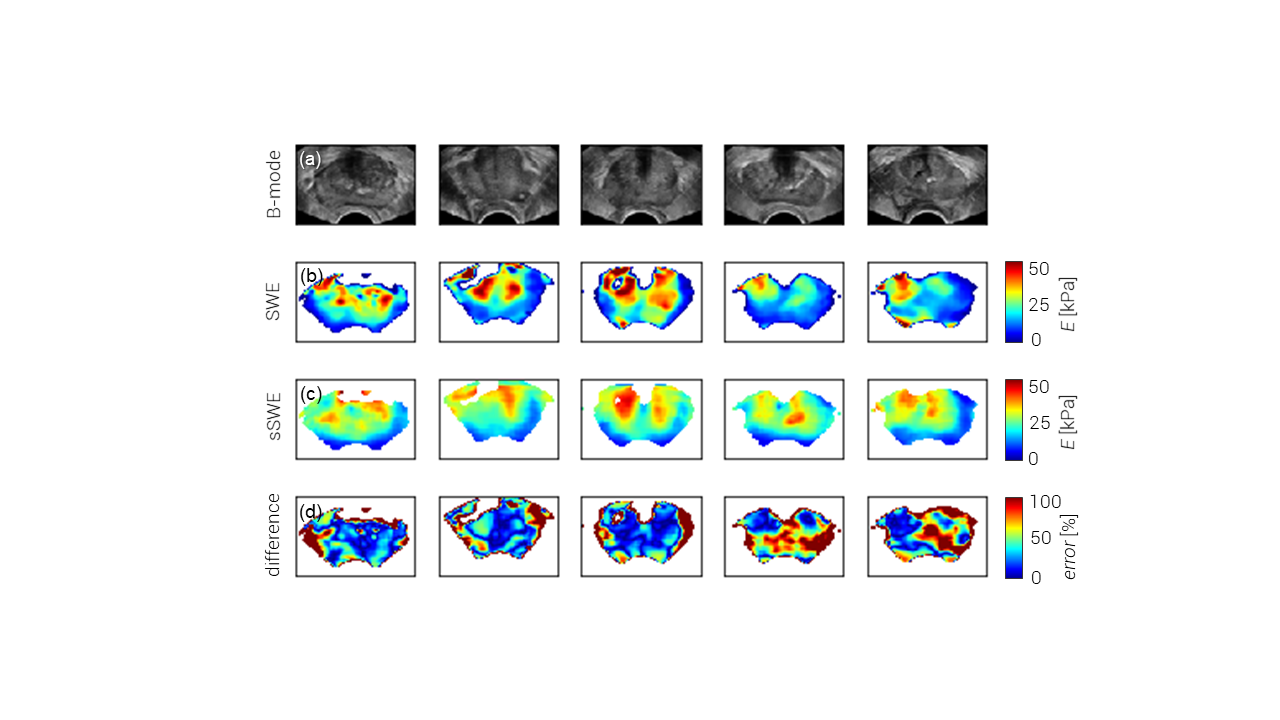}
	\caption{Examples from the ten test patients, with (a) B-mode ultrasound imaging, (b) shear-wave elastographic acquisition, (c) corresponding synthetic SWE (sSWE) image by deep learning, and (d) difference image between sSWE and SWE {showing the error as a percentage of the original sSWE value}.}
	\label{examples2}
\end{figure*}

Every two convolutional layers were followed by a 2$\times$2 spatial max-pooling operation with stride of 2, reducing the image dimensions with a factor 2 and forcing the network to subsequently learn larger-scale features that are less sensitive to local variations. The max pooling operation reduces a kernel of four pixels into one by projecting only the highest value onto the smaller grid \cite{MarcAurelioRanzato}. In total, the encoder consists of 4 convolutional and 2 max-pooling layers mapping the input images into the latent space, which consists of 2 convolutional layers as well. With the decoder being a mirrored version of the encoder layer, appended with a final output layer, the network comprises a total of {158,177} trainable parameters. Max-pooling layers in the decoder are replaced by upsample layers that restore the original image dimensions through nearest-neighbour interpolation. The final output layer consists of a sigmoid activation function that maps the network outputs to the normalized Young's modulus. The use of a sigmoid activation function forces the network to be the most sensitive to values around 0.5 \cite{Han1995a}. Due to the normalization by 100 kPa, the network therefore focuses the most clinically relevant Young's moduli which are in the range between 25 kPa and 75 kPa \cite{Rouviere2017}. The same holds for the {shear-wave speed range of 2.5 to 7.5 m/s within the} thyroid \cite{Zhao2019}.

\subsection{Training Strategy}
Optimization of the trainable DCNN parameters $\theta$ was achieved through minimization of the root-mean-square prediction error (RMSE). Given a set of SWE images $\mathbf{Y}$ and corresponding B-mode images $\mathbf{X}$, we iteratively update the parameters $\theta$ in our network such that the loss of the estimated sSWE images $\mathbf{F}(\mathbf{X}_i;\theta)$ with regard to $\mathbf{Y}$ is minimized:  

\begin{equation}
\label{eqn_RMSE}
L_{RMSE}(\theta)=\sqrt{\frac{1}{N} \sum_{i=1}^{N} |\mathbf{Y}_i-\mathbf{F}(\mathbf{X}_i;\theta)|^2}.
\end{equation}

In this formulation, $N$ is the number of training images. 

Network parameters were learned by employment of the stochastic optimization method Adam \cite{Kingma2014} in 2,100 epochs, using a mini-batch size of 64 training samples for each iteration. {These values were chosen after performing a preliminary {grid-search-based} optimization procedure including the batch size {(i.e., 16, 32, and 64)}, the number of layers {(i.e., 2, 3, and 4 sets of 2-layer blocks in the encoder)}, and the number of epochs {(i.e., 2,100; 2,450; and 2,800)}, using 30 patients from the training set for training and 10 for validation.} {The relatively small batch size is favourable for its looser memory requirements, while preserving an appropriate convergence rate} \cite{Csiba2018}. All filter weights were initialized by a random uniform kernel initializer over the range [-0.05, 0.05] and all biases were initialized to zero. An adaptive learning rate reduction strategy was used to reduce the learning rate once the optimization reached a plateau for 10 epochs.

Whereas B-mode data were available for the full image space, SWE values are only estimated in a certain region of interest. Moreover, SWE analysis allows for a measure of estimation confidence and, usually, low-confidence values are displayed more transparently or not at all. We exploited this information by only propagating loss gradients for those pixels presenting an SWE label of sufficient quality, {using a $>$0.75-confidence threshold as determined by qualitative assessment of the confidence maps}.

Generalizability was promoted through data augmentation, altering {a heuristically chosen} 90\% of the mini-batch data before being fed into the network \cite{Krizhevsky2012}. Data augmentation entailed mirroring and cropping of the image {by maximum 5\% on all sides}, contrast reduction or amplification {with a maximum of 50\%}, random rotation with a maximum of 10 degrees, and full image translation {with a maximum of 50\% laterally and 10\% axially.}. All coordinate transformations were also applied to the SWE labels. Furthermore, we applied drop-out after each max-pooling step to avoid overfitting \cite{Srivastava2014}. This regularization method involves the removal of (in our case 50\% of the) nodes in a random fashion at each training epoch, while switching on all units during testing. As a consequence, inference is based on an approximate average of all these trained dropout networks \cite{Srivastava2014}, acting as an ensemble.

The model was implemented using Keras with the TensorFlow (Google, Mountain View, CA) back-end. Both for training and inference, we employed a Titan XP (NVIDIA, Santa Clara, CA).

\begin{figure*}[]
	\centering
	\includegraphics[trim=2.7in 2.5in 2.1in 1in,clip,width=0.98\textwidth,clip]{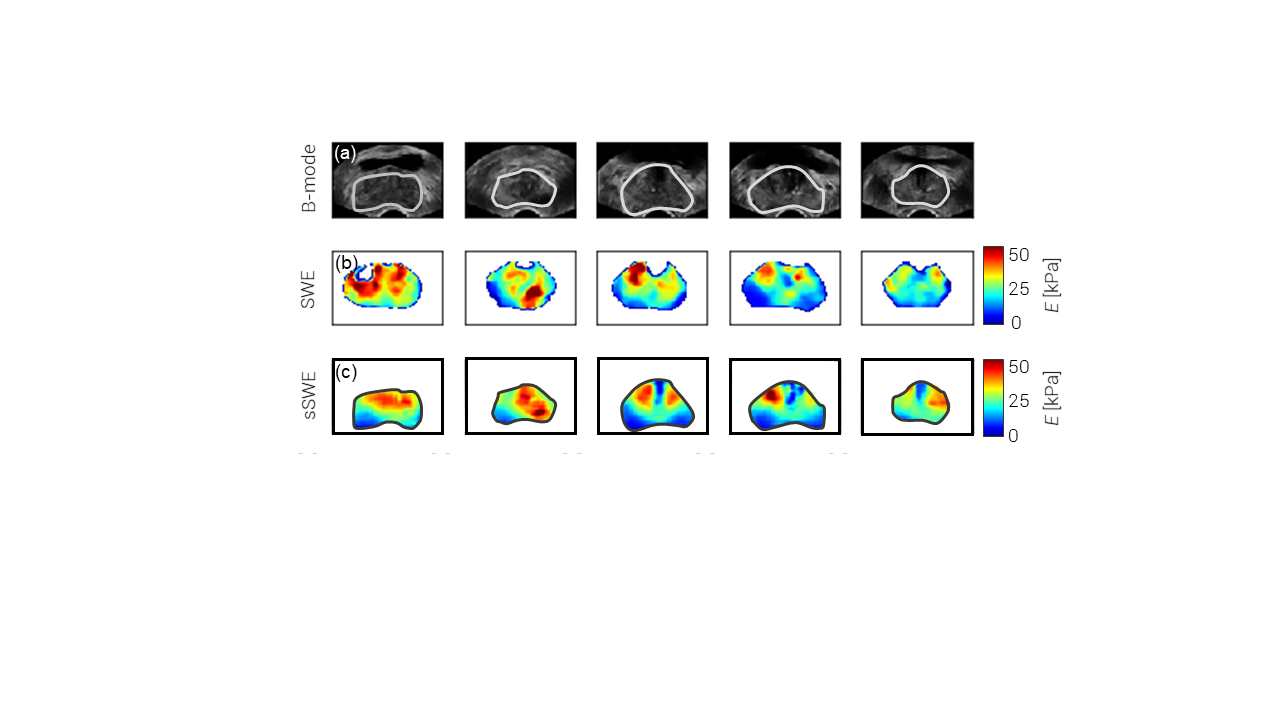}
	\caption{Examples of sSWE generalisation to full-screen B-mode acquisitions in the test patients {depicted in the upper part of Figure 3}, with (a) B-mode ultrasound imaging, (b) corresponding shear-wave elastographic acquisition, and (c) corresponding synthetic SWE (sSWE) image by deep learning.}
	\label{bmodes}
\end{figure*}

\begin{figure}[]
	\centering
	\includegraphics[trim=0.2in 5.7in 8.7in 0.2in,clip,width=0.5\textwidth,clip]{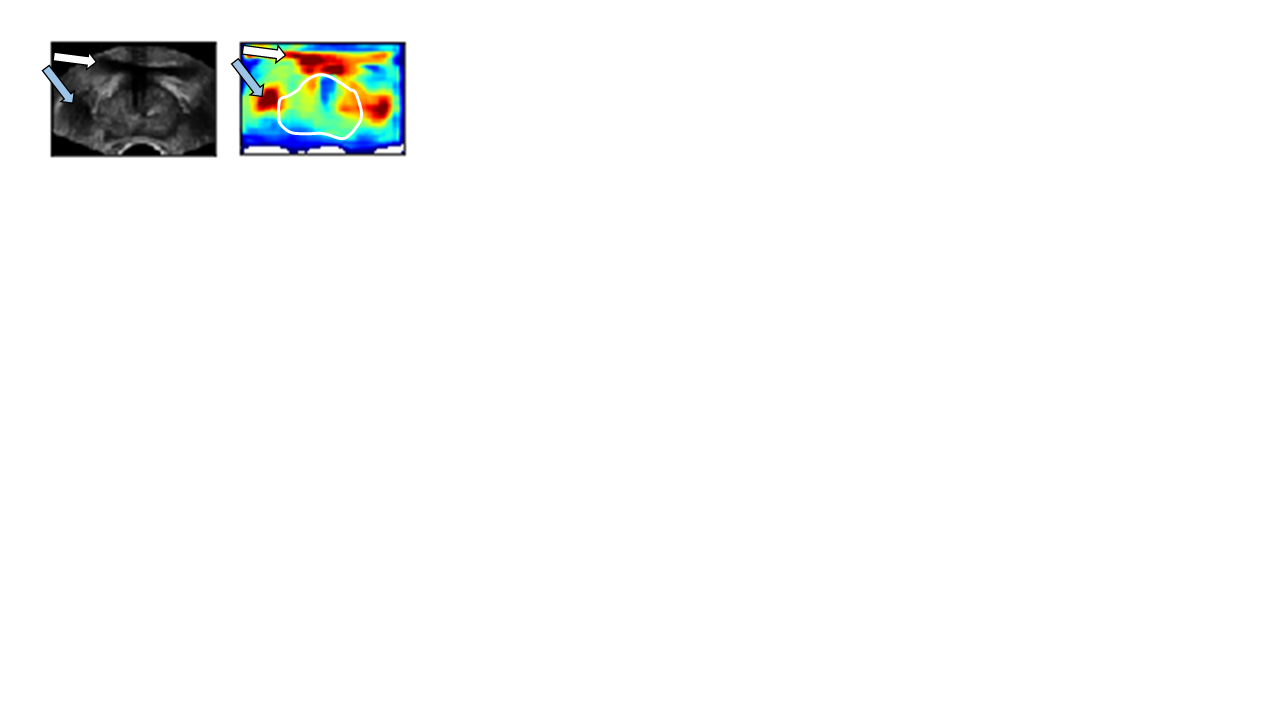}
	\caption{Network output for the full B-mode image of the last prostate in Figure \ref{bmodes}. Extra-prostatic features such as the bladder (indicated by the white arrow) and a vessel structure (indicated by the blue arrow) are mapped as high-elastic areas.}
	\label{outside}
\end{figure}

\subsection{Validation methodology}
Prior to training, our prostate dataset was divided in a training set of 40 patients (consisting of {375} transrectal side-by-side B-mode-SWE images with a varying region-of-interest size, {as the elastogram size was adjusted to fit (half of) the prostate cross-section during the acquisition}) and a test set of 10 patients ({30 images}). All images from the training-set patients were used to maximize the training input and reduce the impact of artefacts, whereas only the three full-prostate images of each test patient were used during testing to ensure that all prostate regions equally contributed to the validation. To evaluate the performance of the DCNN, both the RMSE and mean absolute error (MAE) were monitored:

\begin{equation}
\label{eqn_MAE}
L_{MAE}(\theta)=\frac{1}{N} \sum_{i=1}^{N} |\mathbf{Y}_i-\mathbf{F}(\mathbf{X}_i;\theta)|.
\end{equation}

The RMSE was chosen as loss function because it more heavily penalizes large errors than the similar MAE, and thus allows us to put more weight on the accurate estimation of occasionally-occurring lesions in otherwise low-to-medium-elasticity images. For validation we also considered the mean error (ME),

\begin{equation}
\label{eqn_ME}
L_{ME}(\theta)=\frac{1}{N} \sum_{i=1}^{N} (\mathbf{Y}_i-\mathbf{F}(\mathbf{X}_i;\theta)),
\end{equation}

a measure that reflects a potential bias towards higher or lower Young's moduli.

\begin{figure*}[]
	\centering
	\includegraphics[trim=1.2in 1.5in 0.9in 1.5in,width=0.98\textwidth,clip]{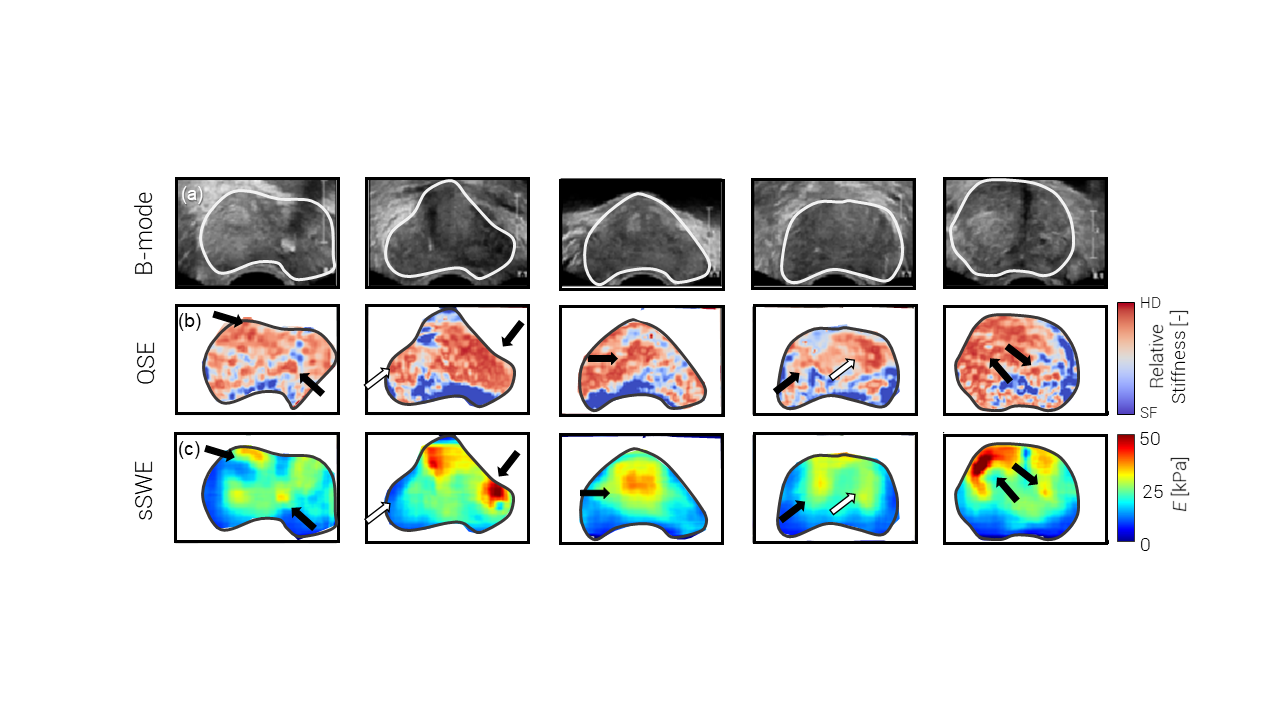}
	\caption{Examples of sSWE results in a non-SWE ultrasound device, with (a) B-mode ultrasound imaging, (b)  quasi-static elastographic acquisition, and (c) corresponding synthetic SWE (sSWE) image by deep learning.}
	\label{other_machine}
\end{figure*}

\begin{figure}[]
	\centering
	\includegraphics[trim=0.6in 0in 0.9in 0in,width=0.5\textwidth]{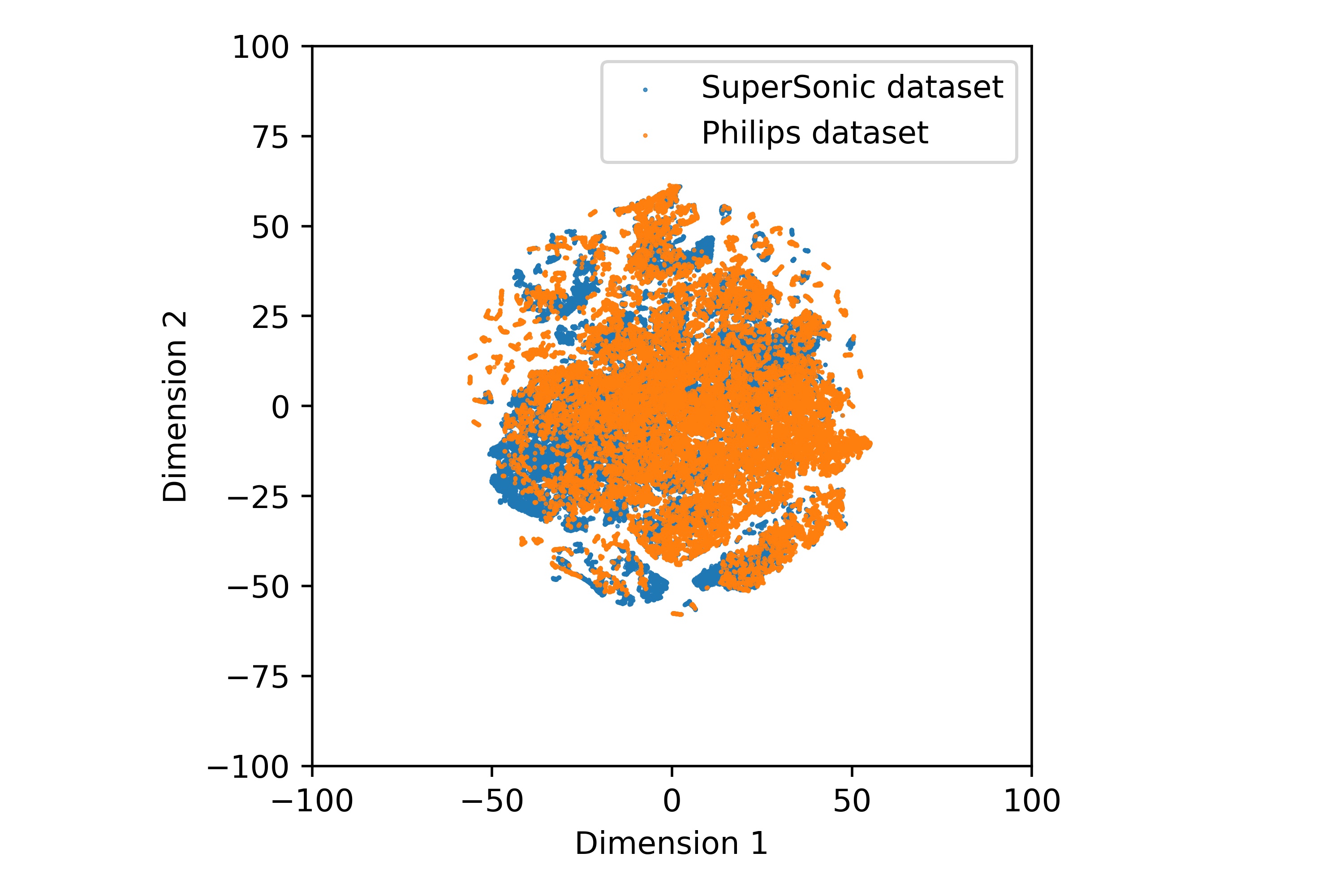}
	\caption{Visualization of B-mode images from both the original SuperSonic Aixplorer ultrasound scanner and the Philips iU22 scanner encoded into high-level features by the DCNN. Reduction of the dimensionality was carried out through t-Distributed Stochastic Neighbor Embedding into two dimensions.}
	\label{t-SNE}
\end{figure}

\begin{figure*}[]
	\centering
	\includegraphics[trim=2.2in 1in 2in 1in,clip,width=0.98\textwidth,clip]{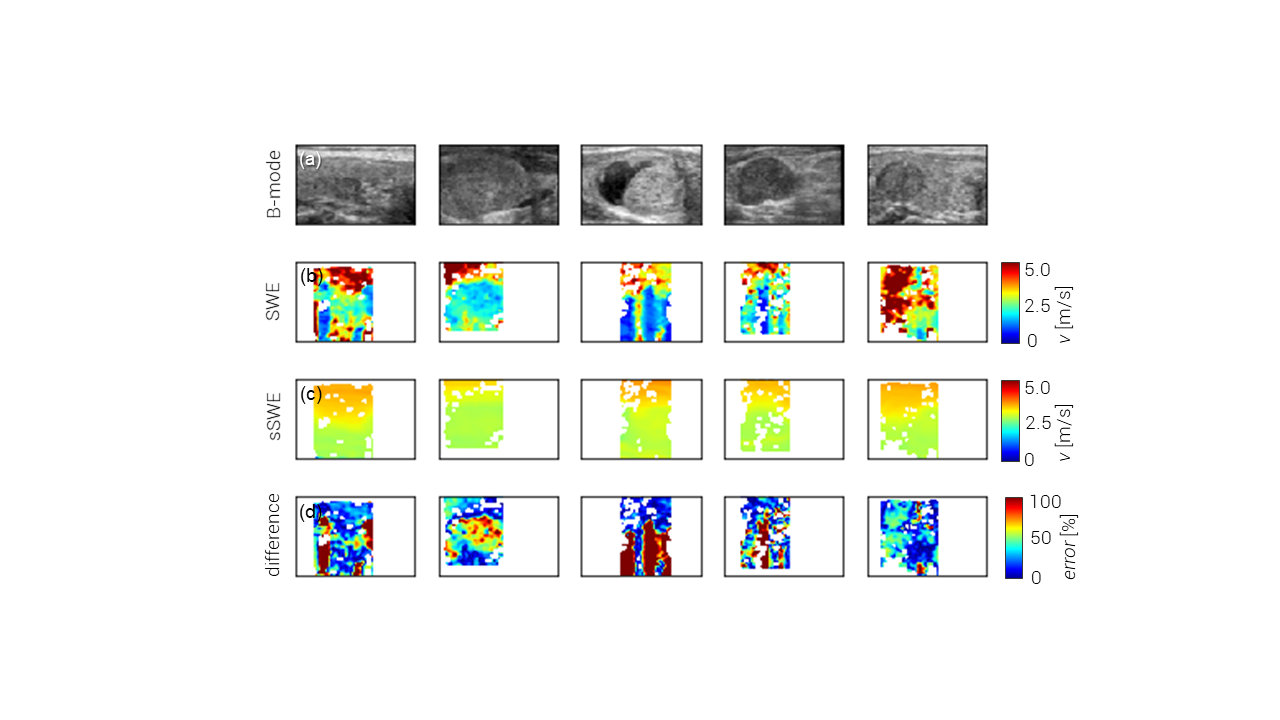}
	\caption{Examples of sSWE of the thyroid, with (a) B-mode ultrasound imaging, (b) corresponding shear-wave elastographic acquisition, (c) corresponding synthetic SWE (sSWE) image by deep learning, and (d) difference image between sSWE and SWE {showing the error as a percentage of the original sSWE value}.}
	\label{thyroid}
\end{figure*}

In order to study to what extent higher-level features are independent from the machine used from the B-mode acquisition, we encoded both the B-mode images recorded with the Philips iU22 scanner and the B-mode images from the test set obtained with the original SuperSonic Aixplorer device. Subsequently, we examined the latent feature space through t-Distributed Stochastic Neighbor Embedding (t-SNE), a probabilistic approach to dimensionality reduction \cite{Maaten2008}.

As tissue structures differ from organ to organ, the thyroid DCNN was separately trained with imaging data of 165 patients and subsequently validated against the test set of the remaining 50 patients. No alterations with regard to the network architecture, processing steps, and training procedure used for the prostate dataset were made, other than the use of the normalised SWE speed as a measure of elasticity. The same performance measures were used.\\

\section{Results}
In Figure~\ref{examples2}, sSWE examples from five test patients are depicted alongside the B-mode and corresponding SWE images. Over the test set, we were able to reach {a per-patient RMSE of 8.8$\pm$1.7~kPa, an ME of -1.6$\pm$1.6~kPa and an MAE of 4.5$\pm$0.96~kPa}. The negative ME reveals that the model is slightly biased towards higher SWE estimates. Qualitatively, tumour locations recognizable on SWE seem to be well estimated also by the sSWE. Outside of the prostate, the SWE as well as the sSWE are generally of lower quality. {The importance of data augmentation is demonstrated by the fact that the performance of the network drops when augmentation is omitted from the procedure, exhibiting an RMSE of 9.9$\pm$2.3~kPa (\textit{p}=0.0035), an ME of -1.7$\pm$1.6~kPa (\textit{p}=0.59) and an MAE of 4.8$\pm$1.2~kPa (\textit{p}=0.026). The reported \textit{p}-values reflect the significance of the improvement as evaluated by a paired \textit{t}-test.} 

Using full-screen B-mode acquisition of the same imaging planes, we demonstrate the ability of sSWE to generalise to B-mode images outside the SWE module. {These B-mode images exhibit a different resolution and contrast compared to the side-by-side B-mode images, and since they were obtained separately, the use of shear waves in the acquisition sequence of side-by-side imaging cannot have played a confounding role}. Nevertheless, even though we allowed the probe to put more pressure on the prostate, generally bringing the prostate closer into view, Figure~\ref{bmodes} shows how the results of these images as input for the trained sSWE model compare well qualitatively to the corresponding SWE images. This suggests that the DCNN extracts higher-level features that are shared among transrectal B-mode images in general. {As the network is only trained on prostate tissue, the output outside of the prostate boundaries is highly variable and remains unvalidated. As highlighted in Figure}~\ref{outside}, {it is unlikely that even tissues close to the prostate are characterized with the same accuracy.}

As can be appreciated in Figure \ref{t-SNE}, depicting the results of t-SNE of the latent feature space, there is only a slight difference in how data from the iU22 and Aixplorer US scanners is mapped into the resulting two-dimensional subspace. {This suggests that the information encoded in the  features that are not specific to the machine with which the data were obtained}. Figure~\ref{other_machine} {depicts B-mode images obtained using a Philips scanner without an SWE option. It demonstrates that some stiff regions as revealed by sSWE correspond to those found by QSE, which was available on the device. The fourth example also shows an example of a false-negative, high-stiffness region that in healthy individuals marks the transition zone of the prostate} \cite{Rouviere2017}{; the second example shows a similar false reading. These are indicated by white arrows.}

Although the prostate-based network could not be applied to generate accurate sSWE images of the thyroid, training of the exact same network architecture with a set of thyroid SWE images resulted in sSWE with {a per-patient RMSE of 0.73$\pm$0.24~m/s, an ME of 0.043$\pm$0.21~m/s and an MAE of 0.34$\pm$0.14~m/s}. {Without retraining the network on thyroid data, we obtained a per-patient RMSE of 1.01$\pm$0.33~m/s, an ME of 0.26$\pm$0.29~m/s and an MAE of 0.46$\pm$0.20~m/s.} {Typical examples of thyroid sSWE images alongside the actual SWE recordings are depicted in Figure} \ref{thyroid}; {sSWE shows general agreement in the differentiation between stiff and soft regions, but the networks capability to show details is limited.} As we did not have access to the SWE confidence, artefacts could not be excluded from training. \\

\section{Discussion}
In this work, we describe and validate a DCNN architecture that provides synthetic SWE images based on B-mode ultrasound. This approach is in line with other recently-proposed inter-modality image synthesis techniques, such as computed tomography from magnetic resonance images \cite{Han2017,Huynh2016,Wolterink2017} or \textit{vice versa} \cite{Jin2018}. Validation in 30 full-prostate SWE images from 10 patients demonstrated a pixel-wise MAE of 4.5$\pm$0.96~kPa, less than 10\% deviation in the clinically-relevant {elasticity} range of 0-70~kPa. Similar results were achieved in the thyroid. Accordingly, it seems that B-mode ultrasound (patterns) harbours information that can be linked tissue elasticity.

{Although the results in this article show the technical feasibility of such an approach, the current study is limited in a few aspects. First of all, the clinical utility of sSWE remains to be investigated. In fact, the use of SWE itself is still being studied in the clinic} \cite{Sang2017,VanHove2014}. {The prostatic SWE images used as input in this study were previously clinically examined for their use in prostate cancer detection, revealing diagnostic potential, especially when used concurrently with other US-based prostate imaging modalities} \cite{Mannaerts2019a}. {The gained experience formed the basis for the qualitative comparison of lesion persistence from SWE to sSWE in this work.}

{A second limitation is in that sSWE does not use mechanical stimulation and can therefore only be considered as a surrogate for elasticity imaging. In this respect, we see sSWE as an elasticity-guided method of tissue typing rather than an alternative to US elasticity imaging techniques. As a consequence of relying solely on B-mode acquisitions, however, sSWE would be less sensitive to e.g. probe pressure, motion artefacts, and region of interest} \cite{Bouchet}. {The clinical interpretability of sSWE, also in relation to SWE, should in the future be examined in a blind fashion.}

{A third limitation resides in the fact that networks have to be retrained when imaging another organ. This emerged by applying sSWE in both the prostate and the thyroid, which are positioned differently and composed of different tissues. On the one hand, this shows that standardization of the imaging procedure is essential for sSWE. Training an sSWE network for organ-specific imaging would therefore be an important part of the standardization procedure. On the other hand, it could indicate that the deep network used in sSWE is to some extent tuned to the tissues that are imaged. In terms of performance, sSWE of the thyroid only seems to capture stiff and elastic regions on a higher scale.} This could be the result of a more diverse range of tissues, a lower degree of standardization in the acquisition procedure and settings, or the absence of means to exclude low-confidence SWE estimation from the analysis.

{It should furthermore be noted that} these results are preliminary in the sense that only a small dataset of a specific organ and a limited number of machines has been taken into account. To provide more robust evidence for the proof-of-principle work presented in this paper, a larger and more variant SWE dataset containing different organs and acquisitions should be examined. The availability of a higher variety of data might also allow the training of a deeper network, which may result in more robust and potentially more accurate sSWE estimation. An in-depth study of SWE images that were incorrectly estimated might guide towards more effective augmentation techniques or highlight the type of acquisitions that should be more abundant in the training set for future data collection.

{In the future, as we already found indications that sSWE might be generalisable to other ultrasound machines, the use of domain adaptation techniques to ensure high-quality, machine-independent sSWE should be investigated} \cite{Wang2018a}. As shown in Figure \ref{t-SNE}, the high-level feature values generally differ little and minimal domain adaptation strategies could already enforce full overlap. For this, for example, shift techniques could be utilized to adjust the mean and variance of the latent throughput. Moreover, the proposed network could possibly be extended with a concurrent estimation of SWE confidence to identify low-confidence regions due to shear-wave artefacts such as signal voids in (pseudo)liquid lesions or B-mode artefacts such as shadowing or reverberation. {Alternatively, an sSWE implementation could be extended to predict other elasticity-related parameters than the Young's modulus or shear-wave speed, such as viscosity} \cite{VanSloun2017}{, which is considered an additional biomarker for cancer in e.g. the prostate} \cite{Zhang2008}. At the present moment, however, there is still a lack of accurate techniques that can assess tissue viscoelastic properties at high spatial resolution allowing the development of such networks.

{However, before using sSWE in the clinic, the clinical potential of the technique for the diagnosis of e.g. prostate cancer should first be investigated. Also its use in registration technology using mechanical properties} \cite{Wildeboer2018d}{, the (automatic) identification of anatomical zones} \cite{VanSloun2019a}{ and a possible role of sSWE features in computer-aided detection approaches could be taken into account} \cite{Lemaitre2015}. {If proven useful, sSWE would be a fast addition to the clinical workflow in situations where conventional SWE is not available or not possible.}

\section{Conclusion}
In conclusion, we have proposed a DCNN architecture that generates synthetic SWE images based on B-mode ultrasound acquisitions. Although further validation of the method is still required, development of this technique paves the way towards elasticity-guided tissue characterisation without the need for complex SWE imaging schemes, {using B-mode characteristics to infer mechanical properties. This would eventually enable SWE-like analysis by basic US scanners}, which could even be low-end systems.

\section{Acknowledgements}
This study has received funding from the Dutch Cancer Society (\#UVA2013-5941) and a European Research Council Starting Grant (\#280209), and was performed within the framework of the IMPULS2-program within the Eindhoven University of Technology in collaboration with Philips.

%\section*{References}
%\biboptions{authoryear}
\bibliography{C:/Users/User/Documents/current_references}
%\bibliography{IEEEabrv,C:/Users/rwildeboer/Documents/current_references}
%\printbibliography

\end{document}